\documentclass[12pt,preprint]{aastex}
\usepackage{subfigure}
\usepackage{epstopdf}
\usepackage{amsmath}
\usepackage{txfonts}
\usepackage{color}

\newcommand{\msol}{\mbox{$\rm{M_{\odot}\ }$}}
\newcommand{\lsol}{\mbox{$\rm{L_{\odot}\ }$}}
\newcommand{\Rsol}{\mbox{$\rm{R_{\odot}\ }$}}
\newcommand{\rsol}{\mbox{$\rm{R_{\odot}\ }$}}
\newcommand{\sol}{\mbox{$\rm{_{\odot}\ }$}}

\begin{document}
\title{The Chemical Composition of $\tau$ Ceti and Possible Effects on Terrestrial Planets }

\author{Michael Pagano\altaffilmark{1}, Amanda Truitt\altaffilmark{1}, Patrick A. Young\altaffilmark{1}, Sang-Heon Shim\altaffilmark{1}}

\altaffiltext{1}{School of Earth and Space Exploration, Arizona State 
University, Tempe, AZ 85287}

\begin{abstract}

$\tau$ Ceti (HD10700), a G8 dwarf with mass 0.78\,M$_\odot$, is a close (3.65 pc) sun-like star where 5 possibly terrestrial planet candidates (minimum masses of 2, 3.1, 3.5, 4.3, and 6.7\,M$_\Earth$) have recently been discovered. We report abundances of 23 elements using spectra from the MIKE spectrograph on Magellan. We find $[\mathrm{Fe}/\mathrm{H}] = -0.49$ and $T_{eff} = 5387$\,K\@. Using stellar models with the abundances determined here, we calculate the position of the classical habitable zone with time. At the current best fit age, $7.63^{+0.87}_{-1.5}$\,Gy, up to two planets (e and f) may be in the habitable zone, depending on atmospheric properties.  The Mg/Si ratio of the star is found to be 2.01, which is much greater than for Earth ($\sim$1.2). With a system that has such an excess of Mg to Si ratio it is possible that the mineralogical make-up of planets around $\tau$ Ceti could be significantly different from that of Earth, with possible oversaturation of MgO, resulting in an increase in the content of olivine and ferropericlase compared with Earth. The increase in MgO would have a drastic impact on the rheology of the mantles of the planets around $\tau$ Ceti.

\end{abstract}


\section{Introduction}
With the number of extrasolar planets increasing rapidly, it appears that Earth-sized planets in their host star's habitable zone (HZ, \citep[e.g.][]{kasting93, kopp13a}) should be numerous \citep{marcy00,gaidos13,dress13,kast13,pet13}. In two decades we have progressed from having no candidate planets to having too many to practically search for detectable biosignatures. A more nuanced analysis than the location of a planet relative to the instantaneous HZ is necessary to choose among candidates. One way of conceptualizing this process is through a ``detectability index" (DI) would account for the ability of a planet to host life, its ability to maintain biosignatures, and the observational techniques that determine what qualifies as a detectable level. The first two terms depend in part upon the characteristics of the host star that determine the position and time evolution of the HZ.
Earth's biosphere took $\sim$ 2 Gy to produce potentially detectable changes in the amount of non-equilibrium species in the atmosphere. Hence a planet that entered the HZ only a few hundred million years ago as the star gradually increased in luminosity may well be habitable and even inhabited, but not have detectable biosignatures. Both the stellar and planetary properties that determine habitability and detectability for terrestrial planets depend fundamentally on chemical composition. 

The determination of elemental abundances has led to the knowledge that every star has its own unique fingerprint. Variations on element to element scales exist between all stars for every element. As a community, we have begun the exploration of extreme cases, especially for the significant rock- forming elements Mg, Si, O and C \citep{dm10,bond10b,unt13}. Since the stellar enrichments are likely primordial, it should lead to planets forming from the same enriched material \citet{santos01,santos03, valfis05, bond06}. 

Although stellar abundances do not correlate directly with planetary abundances, planet assembly simulations have been used to estimate the final planetary compositions \citep{obrien06,bond10b}. \citet{bond10b} found that substantial stellar enrichments could lead to planets with compositions enhanced relative to Fe in particular elements, especially for the 4 major rock-forming elements Si, Mg, C, and O.  This diversity in bulk compositions could lead to variation in mantles and core compositions of terrestrial planets \citep{elk08}.  Recently much attention has been given to carbon-rich systems \citep{bond10b,kuchner05,madhu}, with \citet{bond10b} finding 21 of their 60 stars having C/O ratio above 0.80, believed to be a threshold for carbon and carbide dominated planets, as opposed to solar system-like silicate dominated planets. There are few observations of stars with extreme Mg/Si, with only a handful of stars over Mg/Si$>$1.9.  \citet{bond10b} used initial values for HD177830 (Mg/Si of 1.93) in simulations, finding an olivine-rich planet outside of 0.3 AU, with up to 22$\%$ wt Mg and Mg/Si = 1.71, much greater than that of Earth.

$\tau$ Ceti is of long-standing scientific and popular interest as a nearby (3.65 pc) G8 dwarf. The recent discovery of planet candidates around $\tau$ Ceti \citep{tuo12} gives us an example of a nearby sun-like star with Super-Earth terrestrial planets.  Two of these planets, $\tau$ Ceti e and f, could be in the HZ. These planets are estimated to have masses of $4.29\pm 2.00\,$ and $6.67\pm 3.50 \mathrm{M}_\Earth$ \citep{tuo12}. A high profile and potentially promising system such as $\tau$ Ceti warrants a careful analysis. In this study we evaluate properties of the system, as a simple conceptual example and how it's properties could affect habitability and the DI from a chemical perspective, contributing to a complete and accurate characterization of the system. 


Section~\ref{s.abund} describes new observations from the MIKE spectrograph on Magellan used to derive abundances for 23 elements. Section~\ref{s.mineral} discusses the possible effects of Mg-rich compositions on the internal dynamics and structure of terrestrial planets. New stellar models and HZ predictions using the newly determined abundances are presented in Section~\ref{s.hz}. The possible implications for the habitability of $\tau$ Ceti e and f are discussed in Section~\ref{s.conclusions}.

\section{Abundance Determinations \label{s.abund}}

\subsection{Observations \label{ss.abund.obs}}

The spectrum for $\tau$ Ceti was taken by Paul Butler from the Carnegie Institute of Washington as part of the Magellan Planet Search Program, using the MIKE echelle spectrograph\citep{bern03} on the 6.5 m Magellan II telescope. The spectrum has R~50,000 in the range 4700 \AA\ to 7100 \AA. Only the red chip data was used for abundance determination. Wavelength calibration was carried out by iodine absorption cell\citep{mb92}. The range of S/N for the sample is 70 to 800.  We took the uncalibrated spectra through the abundance finding process, starting with wavelength calibration in IDL and combining all of the orders into a single FITS file. Equivalent widths were measured by hand in IRAF. Stellar parameters were determined using an iterative routine that simultaneously optimizes stellar effective temperature $T_{eff}$, surface gravity $\log(g)$, and micro turbulent velocity, using balance constraints of the abundances derived from neutral (Fe I) and ionized (Fe II) iron. The script started at solar values (effective temperature $T_{eff}$ = 5777 K, surface gravity $\log(g)$ = 4.44, microturbulent velocity $\xi$ = 1 km/s, and [Fe/H] = 0) in the ATLAS9 model atmosphere grid. The stellar $T_{eff}$ and $\xi$ were determined through balancing the reduced equivalent widths and excitation potentials for Fe I lines. This process yielded an [Fe/H] for the Fe I lines. The [Fe/H] from Fe II is primarily influenced by the stellar surface gravity and was forced to match the [Fe/H] from Fe I lines by adjusting the surface gravity in 0.01 dex intervals until the mean abundance of Fe I lines matches that of Fe II lines. Uncertainties in temperature were determined by adjusting the temperature solution until the correlation between the abundances and excitation potential reached a 1$\sigma$ linear correlation coefficient for the given number of lines. This was done by slowly increasing the temperature until there was a linear 1$\sigma$ deviation from the calculated value. The same 1$\sigma$ approach was used for the $\xi$ uncertainty but with the abundances versus reduced equivalent widths. The errors for [Fe/H] were determined by using a standard deviation of all the line abundances for Fe I, combined with the change resulting from varying the Tspec and $\xi$ by their 1$\sigma$ uncertainties. The elemental abundances were determined using the spectral analysis program MOOG \citep{sneden73}, with a model atmosphere created by the program mspawn through ATLAS9 \citep{kurucz93}. The errors for these elements are determined from the standard deviation of abundances given for each spectral line. \textbf{At such low metallicities, the measurement of oxygen for the weak lines found in this spectrum becomes difficult, but remain important. Therefore,  the errors for oxygen were obtained using methods from \citet{lis15}. These are a combination of the differences in abundances between the spectral lines, by varying the stellar parameters by the parameter errors( 53 K, .06 dex surface gravity, 0.08 dex metallicity, and 0.23 $kms^-1$ microturbulent velocity) to account for the sensitivity to these parameters, and by uncertainties in the equivalent widths due to continuum placement.  Since the oxygen lines at this low of a metallicity have such shallow spectral features, the placement of the continuum yields the largest source of error.}
\subsection{Abundances \label{ss.abund.result}}

We find $T_{eff} = 5373$K $\pm$ 53 K and $\log$ (g) = 4.55. $\tau$ Ceti is relatively metal-poor, with $[\mathrm{Fe}/\mathrm{H}] = -0.49 \pm 0.08$. Table~\ref{tab.abund} gives stellar parameters and abundances [X/H]. Values for the absolute number ratios C/O and Mg/Si are also provided. These ratios are taken before normalizing to any solar spectrum. Due to the low metallicity and the spectral range, the C/O ratio has a fairly large uncertainty. Table ~\ref{tab.list} gives the line list and line parameters. The solar values were calculated using a solar spectrum and compared to Tau Ceti on a line to line basis to calculate the solar normalized values in ~\ref{tab.abund}, but no solar values were used when dealing with elemental ratios such as Mg/Si.

This work yields Mg/Si = $1.78\pm0.04$, much higher than Earth Mg/Si = 1.21 \citep{kargal93} and the sun's Mg/Si = 1.05 \citep{asplund05}. The absolute abundance of Mg is higher than the sun, despite the star's low metallicity. Ca is also supersolar. The Al/Fe is enhanced, but the absolute abundance of Al is slightly less than solar. 

The values of Mg/Si in the literature vary considerably. Tau Ceti appears to have a high Mg/Si ratio compared to the sun (1.78 in this study, with a mean of 1.48 in literature sources\citep{thev98, bond06, bond08, take07, neves09, adi12}. The values range from 1.29\citep{adi12} to 1.78. However, since the Mg/Si ratio is dependent on 2 separate unnormalized values, small deviations in both values can lead to larger differences in the ratio. It can be more useful to check the Mg and Si values individually to see the true variation in the Mg/Si ratio. The absolute (not solar normalized) values for Si from 9 different studies, including this one\citep{thev98,bod03,valfis05,gilli06,bond06,take07,neves09,adi12}  range from 7.12\citep{gilli06} to 7.33 \citep{thev98}, with a mean of 7.22 and median of 7.2. The value reported here is exactly the median of 7.2. The Mg values from 6 different studies, including this one\citep{thev98,take07,neves09,milone11,adi12} range from 7.3 to 7.53 with a mean of 7.41 and a median of 7.43. This study finds a Mg/H value of 7.45, which once again is close to the median value.  Varying the our values by the errors from this study ($\pm$0.05 for Mg and $\pm$0.04 for Si) the value for Mg/Si can be decreased to as low as 1.33 or as high as 2.19. Both this work and the median values of other careful determinations in the literature point to a high Mg/Si ratio for $\tau$ Ceti, and even the lowest reported values are still substantially higher than the sun. Even if $\tau$ Ceti eventually proves to be at the low end of this range, its example indicates that the possibility of exoplanets with exotic silicate mineralogy needs to be explored. 

\begin{table}
\caption{Abundances of $\tau$ Ceti}
\label{tab.abund}
\scriptsize
\begin{center}
\begin{tabular}{ccc}
\hline
\hline
 & Value & {$\pm$}\\
\hline
Temp & 5373 & 53 \\
MtVel & 1.30 & 0.23 \\
log(g) & 4.55 & 0.06 \\
$[$Fe/H$]$ & -0.49 & 0.08 \\
$[$C/H$]$ & -0.84 & 0.14 \\
$[$O/H$]$ & -0.48 & 0.12 \\
$[$Na/H$]$ & -0.24 & 0.06 \\
$[$Mg/H$]$ & 0.14 & 0.03 \\
$[$Al/H$]$ & -0.07 & 0.01 \\
$[$Si/H$]$ & -0.28 & 0.04 \\
$[$S/H$]$ & -0.49 & 0.11 \\
$[$Ca/H$]$ & 0.04 & 0.10 \\
$[$Sc/H$]$ & -0.32 & 0.09 \\
$[$Ti/H$]$ & 0.16 & 0.08 \\
$[$V/H$]$ & 0.16 & 0.05 \\
$[$Cr/H$]$ & -0.33 & 0.22 \\
$[$Mn/H$]$ & -0.26 & 0.18 \\
$[$Co/H$]$ & -0.21 & 0.04 \\
$[$Ni/H$]$ & -0.21 & 0.12 \\
$[$Cu/H$]$ & -0.20 & 0.15 \\
$[$Zn/H$]$ & -0.83 & * \\
$[$Y/H$]$ & -0.26 & 0.30 \\
$[$Mo/H$]$ & -0.18 & * \\
$[$Ba/H$]$ & -0.44 & 0.02 \\
$[$La/H$]$ & -0.16 & * \\
$[$Ce/H$]$ & 0.23 & * \\
$[$Nd/H$]$ & -0.03 & 0.38 \\
$[$Eu/H$]$ & -0.11 & * \\
C/O & 0.27 & 0.21 \\
Mg/Si & 1.78 & 0.06 \\
\hline
\end{tabular}
\end{center}
\end{table}

\begin{table}
\caption{Complete line list \tablenotemark{a}}
\label{tab.list}
\scriptsize
\begin{center}
\begin{tabular}{cccccc}
\hline
\hline
 Line (\AA\ & Element & Lower Excitation Potential & log (Oscillator Strength) & EQW solar & EQW Tau Ceti\\
\hline
5052.15	&	C	&	7.68	&	-1.3	&	37.8	&	8.29	\\
5380.32	&	C	&	7.68	&	-1.61	&	21.9	&	4.65	\\
6156.8	&	O	&	10.74	&	-0.43	&	4.1	&	1.54	\\
6363.79	&	O	&	0	&	-9.72	&	4.9	&	1.65	\\
6154.23	&	Na	&	2.1	&	-1.53	&	39.8	&	23.91	\\
6160.75	&	Na	&	2.1	&	-1.23	&	58.4	&	43.52	\\
5711.09 & Mg & 4.35 & -1.83 & 104.1 & 113.7 \\
6318.72 & Mg & 5.11 & -1.99 & 38.2 & 51.6 \\
6696.03	&	Al	&	3.14	&	-1.58	&	38.1	&	34.94	\\
6698.67	&	Al	&	3.14	&	-1.95	&	21.9	&	19.3	\\
5690.43 & Si & 4.93 & -1.77 & 52.6 & 37.24 \\
5793.08 & Si & 4.93 & -2.06 & 48.2 & 28.3 \\
6125.03 & Si & 5.61 & -1.51 & 34 & 18.67 \\
6142.48 & Si & 5.62 & -1.54 & 36.3 & 23.49 \\
6145.01 & Si & 5.62 & -1.36 & 41.6 & 23.68 \\
6155.13 & Si & 5.62 & -0.78 & 81.4 & 55.65 \\
6244.48 & Si & 5.61 & -1.36 & 42.4 & 28.69 \\
6721.86 & Si & 5.86 & -0.94 & 47.8 & 28.65 \\
\hline
\end{tabular}
\end{center}
\tablenotetext{a}{Table 2 is published in its entirety in the electronic edition of ApJ, A portion is shown here for guidance regarding its form and content.}
\end{table}

\section{Internal Structure and Dynamics of Mg-rich Terrestrial Planets \label{s.mineral}}

Though the range of Mg and Si from different groups vary more than quoted errors for this star and many others\citep{hinkel14}, $\tau$ CetiÕs Mg/Si is significantly higher than the solar value of 1.05 (Asplund 2005). Regardless of the precise value, the example of $\tau$ Ceti encourages us to explore the geophysical effects of distinctly non-terrestrial compositions on extrasolar rocky worlds. The theoretical behaviors of high Mg/Si planets are not based on specific cutoffs, but rather a gradual change from Earth-like to Mg-rich.

The estimated sizes of the candidate exoplanets span a considerable range, from 2.0 to 6.6 $M_{\Earth}$. It is considered likely that they qualify as terrestrial planets, but it is uncertain without an estimate of bulk densities. The following analysis assumes that the planets are terrestrial Super-Earths or that smaller, undiscovered terrestrial planets are present. It also assumes that Mg/Si in terrestrial planets scales roughly with Mg/Si in the star. Stellar abundances most likely represent primordial composition\citep{santos01,santos03, santos05, valfis05, bond06}. We know from the example of Earth that there is not a 1:1 correspondence between the abundance ratios of star and planet (see Section~\ref{ss.abund.result}), but for our solar system at least, the numbers are similar, with the terrestrial value being higher than the solar. This work speculates on the effects of an extreme Mg/Si ratio such as that found for $\tau$ Ceti by some groups, including this study, on terrestrial planets. This kind of analysis would benefit from tighter constraints on the planetary compositions, but stellar properties can provide useful boundary conditions. Such constraints would likely come from models of planet assembly from a protoplanetary disk of $\tau$ Ceti's primordial composition. The compositions of planets can differ from a star's due to various processes occurring during planet formation and evolution.  For example, the Mg/Si ratio of Earth's upper mantle is greater by $\sim$20\% than that of our Sun's photosphere. 
\citet{bond10a} indicates that stellar enhancements will be reflected in planets, even if the exact ratio is not the same. While solubility of Mg may be very small for the metallic iron core and some amount of Si can exist as an impurity in the core, most  Mg and Si may remain in the rocky mantle.

The mantles of planets with Mg/Si approaching 2 will be almost entirely made of olivine, (Mg,Fe)$_2$SiO$_4$, with very little or no pyroxene, (Mg,Fe)(Al,Si)O$_3$.  This is in sharp contrast with the mineralogy of Earth's mantle where the proportions of these two minerals are similar to each other.  Furthermore, the Mg-rich mantle could even contain ferropericlase, (Mg,Fe)O, at shallow depths. Ferropericlase exists only in the lower mantle (700--3000\,km depths) in Earth.  If the Mg-rich terrestrial exoplanets have surface volcanism through (partial) melting of the deep mantle similar to Earth, such as some hot spots and mid oceanic ridges, the high Mg/Si ratio will result in very different magnitude of melting and composition of melts. Mantle melting processes are believed to be responsible for the formation of the crust on Earth.

If the composition remains similar throughout the mantle, the deeper mantle may be composed mainly of Mg-silicate perovskite (Mg,Fe)(Al,Si)O$_3$ and ferropericlase  in similar proportion, unlike the Earth's lower mantle where Mg-silicate perovskite is dominant (70\%).  It has been predicted that Mg-silicate perovskite may break down into oxides (ferropericlase and silica) at 1\,TPa \citep{um06}.  However, the known terrestrial planet candidates around $\tau$ Ceti may not have sufficient mass to reach this pressure.  Therefore, except for the shallow depths ($<600$\,km), the mantles of these large terrestrial planet candidates should be composed mainly of the phases of the Earth's lower mantle but with different proportions.  

The increased amount of ferropericlase may have a profound impact on mantle dynamics.  For the lower mantle, an increase in ferropericlase expected from the elevated Mg/Si ratio ($\simeq$ 2) could decrease the viscosity of the lower mantle by two orders of magnitude \citep{amm11}.  The severe decrease in mantle viscosity could result in more vigorous convection, potentially affecting surface tectonics.  Even if improved determinations of Mg/Si in $\tau$ Ceti converge on a lower value than that found in this study, such novel geophysics are likely to appear in other exoplanet systems.

\section{Habitable Zone \label{s.hz}}
We evaluate the evolution of HZ boundaries defined by theoretical stellar evolution tracks coupled with the radiative-convective atmospheric models of \citet{kopp13a,kopp13b}. 
Evolutionary tracks were created with the TYCHO code \citep{young_2005_aa}. TYCHO is a 1D stellar evolution code with a hydrodynamic formulation of the stellar evolution equations.  It uses OPAL opacities \citep{ir96,alex94,rn02}, a combined OPAL and Timmes equation of state (HELMHOLTZ) \citep{timmes_1999_aa,rn02}, gravitational settling (diffusion) \citep{thoul_1994_aa}, 
automatic rezoning, and an adaptable nuclear reaction network with a sparse solver. A 177 element network terminating at $^{74}$Ge is used throughout the evolution. The network uses the latest REACLIB rates \citep{rauscher_2000_aa, ang99,iliadis_2001_aa,wiescher_2006_aa}, weak rates from \citet{LMP00}, and screening from \citet{graboske_1973_aa}. Neutrino cooling from plasma processes and the Urca process is included. Mass loss is included but is trivial for a 1\msol main sequence star. (Heightened early mass loss seen in some young stars \citep{wood05} is not included. This may play a role in early HZ evolution and be relevant to the ``faint young sun'' paradox). It incorporates a description of turbulent convection \citep{MA07_conv, AMY09, amy10, AM11} which is based on three dimensional, well-resolved simulations of convection sandwiched between stable layers, which were analyzed in detail using a Reynolds decomposition into average and fluctuating quantities.  It has no free parameters to adjust, unlike mixing-length theory (MLT). 

We used an initial mass of 0.784 \msol \citep{tang11}. The initial composition has the elemental ratios derived in Section~\ref{ss.abund.result}, scaled so that when diffusion is included the modeled surface composition matches the measured abundances at the best-fit model age. Tracks were computed to the end of its main sequence lifetime.
In order to find the best-fit age, we step through the evolutionary track and evaluate: 
\begin{equation}
\chi ^2 = \frac{(R_{mod}  R_{obs})^2}{\sigma_R^2} + \frac{(L_{mod}  L_{obs})^2}{\sigma_L^2}
\end{equation}
where $R_{mod}$ and $R_{obs}$ are the model and observed radii, $L_{mod}$ and $L_{obs}$  are the model and observed luminosities, and $\sigma_R$ and $\sigma_L$ are the observational errors in radius and luminosity. The uncertainty in age is given by the ages at which the evolutionary track crosses the observational error box. The adopted radius is 0.7817\rsol, which is the mean of the interferometric and asteroseismological radii (0.773$\pm$0.024\Rsol and 0.07916$\pm$0.016\Rsol, respectively) determined by \citep{tang11}. For our error in $R$ we use the error of the interferometrically defined radius, which encompasses the range of uncertainty of the asteroseismological value. This provides us with a conservative estimate of the age uncertainty. We adopt $L = 0.504$\lsol \citep{tex09,pij03a,pij03b}.  \citet{tang11} find $\log g \simeq 4.53$ using their asteroseismological analysis, similar to our spectroscopically-determined $\log g = 4.55$. The age of the system from this analysis is $7.63^{+0.87}_{-1.5}$Gy.

We evaluate the classical HZ around $\tau$ Ceti, where liquid water is stable at the planetary surface given reasonable assumptions about the planetary atmosphere. 
Using the TYCHO simulation, we can predict the location of the inner and outer edges of the HZ as a function of time, and ask the question of whether a planet would be in the HZ long enough to harbor potential life. Locations for the inner and outer edges of the HZ are calculated using the stellar luminosity and temperature parameterizations from \citet{kopp13a, kopp13b}.  \citet{kopp13a} provide updated 1-D radiative-convective climate models (originally from \citet{kasting93}) to obtain new estimates for HZ widths around F, G, K, and M-type stars. 
Rayleigh scattering by water vapor in the atmosphere of exoplanets with new scattering coefficients is included. The models assume Earth-mass planets with water- (corresponding to the inner HZ edge) or CO$_2$ (corresponding to the outer HZ edge) dominated atmosphere. Both inner and outer edge calculations are based on "inverse climate modeling" where the planetary surface temperature is specified, and the models are used to determine the stellar flux needed to sustain that given temperature. 
One potential issue is that the $\tau$ Ceti planet candidates are all super-Earths. Models of planets with higher masses than earth and thicker atmospheres are discussed, but parameterizations are not provided. For the cases discussed, the edge of the HZ moves by at most a few percent in radius, so for the present we adopt the \citet{kopp13a} parameterizations.

\begin{figure}[h]
\begin{center}
\includegraphics[width=5.0in]{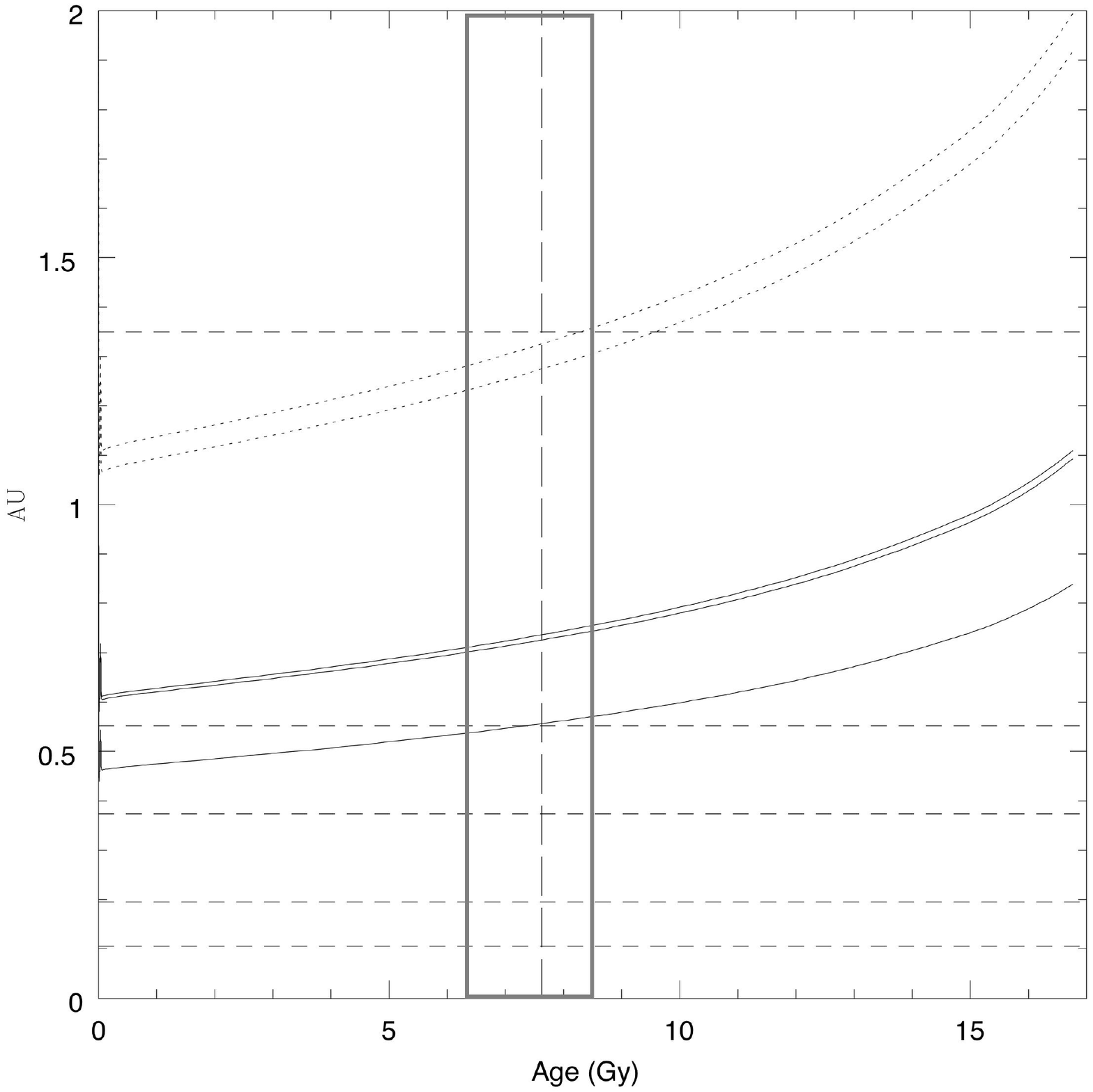} 
\end{center}
\caption{Inner and outer radii of the HZ of $\tau$ Ceti. Solid lines show the inner edge of the HZ for, in order of distance from the star, the Recent Venus, Runaway Greenhouse, and Moist Greenhouse cases. The outer edge is shown by the dotted lines for the Maximum Greenhouse and Early Mars cases. The orbits of the putative planets are indicated by the horizontal dashed lines. The vertical line is the best-fit age of the system at 7.63 Gy with the age uncertainty indicated by the gray box. \label{hzfig}}
\end{figure}

Figure~\ref{hzfig} shows the inner and outer radii of the HZ for the cases in Table~3 of \citet{kopp13a, kopp13b}. 
The orbits of the putative planets are indicated by the horizontal dashed lines. The vertical line is the best-fit age of the system at 7.63\,Gy with the age uncertainty indicated by the gray box. The solid lines show the inner edge of the HZ for, in order of distance from the star, the Recent Venus, Runaway Greenhouse, and Moist Greenhouse cases. The outer edge is shown by the dotted lines for the Maximum Greenhouse and Early Mars cases. On the main sequence the radii of the HZ limits increase monotonically with time as the stellar luminosity increases. 
The three inner candidates are interior to the HZ even for the most optimistic prediction. Candidates e and f are potentially interesting marginal cases. For the Recent Venus case (assuming that Venus moved out of its HZ 1\,Gy ago) planet e is reaching the end of its habitable lifetime. At the best-fit age of $\tau$ Ceti it is right at the edge of the HZ. Future direct observations from space-based interferometers/occulters of this world have the potential to place constraints on climate models. For the theoretically-motivated greenhouse cases, candidate e is always interior to the HZ. Candidate f is potentially more promising, at least judging by the theoretical climate models. The outer edge of the HZ is approaching the position of planet f. Given the uncertainty in the stellar age, arising largely from the observational uncertainty in the star's luminosity, candidate f is between 0 and 1.5\,Gy from entering the HZ defined by the Early Mars case (assuming Mars was habitable 3.8\,Gy ago). The range is extended by $\sim$1\,Gy for the Maximum Greenhouse case. Even in the most pessimistic case, the planet will have about  7\,Gy of habitable lifetime until the end of the main sequence, plus additional time while the star traverses the subgiant branch. From a detectability standpoint, however, f is a poor candidate. At best, the planet has been in the HZ for $<<$1Gy under these assumptions. The rate of change of $L$ and $T_{eff}$ as a function of time  means that cases similar to f where a planet enters the HZ in the latter part of the star's life are more common than planets that have been in the HZ since early times. 
This serves as a reminder that the present ``habitability'' of a planet does not necessarily indicate that it is a good candidate for detecting biosignatures. The temporal evolution of the system must be taken into account.

The general characteristics of the $\tau$ Ceti HZ are also interesting in the event that other planets, perhaps closer to the mass of Earth, are discovered. Table~\ref{habtab} summarizes the range of orbits with given durations of habitability, assuming a stellar age of 7.63\,Gy. We find ranges for orbits that are continuously habitable for the entire main sequence, orbits that remain habitable for at least 4\,Gy (the length of time required for metazoan life to develop on earth), and orbits that are currently habitable and have been so for at least 2\,Gy.  This last case reflects the approximate amount of time taken for life to make potentially detectable changes in the composition of Earth's atmosphere (in particular the presence of free oxygen, methane, and nitrous oxide). Two cases are tabulated, a conservative case for the minimum size HZ using the Moist Greenhouse and Maximum Greenhouse prescriptions, and an optimistic case using the Recent Venus and Early Mars climate models. For the conservative case there are {\it no} continuously habitable orbits. In this dire situation the maximum duration of habitability is a mere 16\,Gy, since a 0.78\sol star has a very long lifetime. 
Of the known candidates, the best case for a planet which has detectable atmospheric signatures of an active biota is $\tau$ Ceti e. The most favorable location for a hypothetical sixth planet that has developed atmospheric biosignatures is the 0.45\,AU range between 0.73 and 1.18\,AU. 

\begin{table}
\caption{Habitable Zone Properties}
\label{habtab} 
\begin{center}
\begin{tabular}{lcc}
\hline
\hline
&Conservative & Optimistic\\
& \multicolumn{2}{c}{AU}\\
\hline
Currently Habitable & 0.73 - 1.28 & 0.55 - 1.32 \\
Continuously Habitable Range & & 0.84 - 1.1 \\
4 Gy Habitable Range & 0.67 - 1.52 & 0.5 - 1.6 \\
Habitable $\le -2$Gy to Present & 0.73 - 1.18 & 0.55 - 1.25 \\
\end{tabular}
\end{center}
\end{table}

\section{Conclusions \label{s.conclusions}}

The $\tau$ Ceti system has been the object of speculation about extraterrestrial life in both the popular imagination and the scientific literature for decades due to its proximity to Earth and the star's sun-like characteristics. The recent discovery of Super-Earth mass planet candidates has further increased interest in the system. Two of the candidates, $\tau$ Ceti e and f, are near the edges of the theoretical HZ. These worlds could potentially be high priority candidates for observations by a future direct detection mission looking for biosignatures. A more thorough analysis of the system is therefore warranted. We specifically examine the contributions of chemical composition to the likelihood of e and f being habitable and having detectable biosignatures. We examine stellar models with the measured composition and the accompanying habitable zone evolution. We also discuss the geophysical implications of a high Mg/Si in planets. Although Mg/Si in $\tau$ Ceti needs to be further constrained, it is clearly larger than in the Sun and Earth. 

Two main considerations turn out to be important for the case of $\tau$ Ceti. First, the star has a very high Mg/Si=1.78. The Mg/Si ratio is closer to the boundary at which a planetary mantle would be expected to transition from a olivine/pyroxene mineralogy to an olivine- dominated mineralogy at shallow depths and a ferropericlase-rich lower mantle. The viscosity of the mantle could be much lower than earth. Coupled with an enhanced abundance of actinide elements contributing to radiogenic heating of the interior, planets around $\tau$ Ceti may have much more vigorous and long-lived mantle convection, which would in turn affect the activity of surface processes and (bio)geochemical cycling. Furthermore, the surface rock composition may be very different from that of Earth as the crust may be formed by partial melting of the mantle. This could be an essential point when considering the habitability and detectability of a planet 7.6 Gy old. While this could benefit the habitability of $\tau$ Ceti's planets, it could interfere with the DI of a younger planet. Faster geochemical cycling could impede the buildup of biologically produced non-equilibrium chemical species in the planet's atmosphere. Time dependent analyses of planetary evolution are required.

Second, the time and composition dependent modeling of the star provides a better estimate of not only the planets' habitability potential now, but their habitability history. Given the HZ models used herein, planet e is very near the edge of the HZ for very optimistic assumptions. For more conservative cases it lies interior to the HZ  at all times. Planet f highlights the time-dependent nature of detectability. The planet is near the edge of the predicted HZ. For the most optimistic case it has only recently entered, while for conservative cases it is still just outside. Even though it will likely see more than 7 Gy of continued habitability as the star ages, as far as we can trust existing climate models it has been in the HZ at most much less than 1 Gy. Earth required $\sim$2 Gy before the biota modified the atmosphere sufficiently to possibly detectable from another stellar system. Even if f's habitability potential is high, its detectability index is low.

In conclusion, the known candidate planets around $\tau$ Ceti are not as promising targets for finding life as they appear at first analysis. They do, however, show the power of full characterization of exoplanetary systems, in particular their chemical composition, for choosing targets for future missions to detect life.


\begin{thebibliography}

\bibitem[Adibekyan et al.(2012)]{adi12} Adibekyan, V. Zh., Sousa, S.G., Santos, N.C., Delgado Mena, E., Gonzalez Hernandez, J.I., Israelian, G., Mayor, M., \& Khachatryan, G. 2012, A\&A 544, 44

\bibitem[Alexander \& Ferguson(1994)]{alex94} Alexander, D.R. and Ferguson, J.W. 1994, \apj,  437, 879

\bibitem[Ammann et al.(2011)]{amm11}Ammann, M.W., Brodholt, J.P., Dobson, D.P.  2011, E\&PSL, 302, 393

\bibitem[Angulo et al.(1999)]{ang99}Angulo, C., Arnould, M., Rayet, M., Descouvemont, P., Baye, D., Leclercq-Willain, C., Coc, A., Barhoumi, S., Aguer, P., Rolfs, C., Kunz, R., Hammer, J. W., Mayer, A., Paradellis, T., Kossionides, S., Chronidou, C., Spyrou, K., DeglInnocenti, S., Fiorentini, G., Ricci, B., Zavatarelli, S., Providencia, C., Wolters, H., Soares, J., Grama, C., Rahighi, J., Shotter, A., and Lamehi Rachti, M. 1999, Nucl. Phys. A, 656, 3

\bibitem[Arnett, Meakin \& Young(2010)]{amy10}Arnett, D., Meakin, C., \& Young, P. 2010, ApJ, 710, 1619

\bibitem[Arnett, Meakin, \& Young(2009)]{AMY09} Arnett, D., Meakin, C., and Young, P.A. 2009, \apj, 690, 1715

\bibitem[Arnett \& Meakin(2011)]{AM11} Arnett, D. and Meakin, C. 2011, \apj, 733, 78

\bibitem[Asplund et al.(2005)]{asplund05} Asplund, M., Grevesse, N., \& Sauval, A. J. 2005, in ASP Conf. Ser. 336, Cosmic Abundances as Records of Stellar Evolution and Nucleosynthesis, ed. T. G. Barnes, III \& F. N. Bash (San Francisco, CA: ASP), 25

\bibitem[Barbuy et al.(2011)]{barbury11} B. Barbuy, M. Spite, V. Hill, F. Primas, B. Plez, R. Cayrel, F. Spite, S. Wanajo, C. Siqueira Mello Jr., J. Andersen, B. Nordstršm, T. C. Beers, P. Bonifacio, P. Franois, and P. Molaro 2011, A\&A, 534, 60

\bibitem[Batalha et al.(2013)]{bat13} Batalha et al. 2013, ApJS 204, 24B

\bibitem[Bernstein et al.(2003)]{bern03} Bernstein, R., Shectman, S.A., Gunnel, S.M., Mocknacki, S., Athey, A.E. 2003, SPIE, 4841, 1694

\bibitem[Beers \& Christlieb(2005)]{beers05}Beers, T.C., \& Christlieb, N. 2005, ARA\&A, 43, 531

\bibitem[Bodaghee et al.(2003)]{bod03} Bodaghee, A., Santos, N.C., Israelian, G., Mayor, M. 2003, A\&A 404, 715

\bibitem[Bond et al.(2006)]{bond06} Bond, J.C. et al. 2006, MNRAS, 370, 163

\bibitem[Bond et al.(2008)]{bond08} Bond, J.C. et al. 2008, ApJ, 682, 1234

\bibitem[Bond et al.(2010a)]{bond10a} Bond, J.C., Lauretta, D.S., OÕBrien, D.P. 2010a, Icarus, 205, 321

\bibitem[Bond et al.(2010b)]{bond10b} Bond, J.C., OÕBrien, D.P., Lauretta, D.S. 2010b, ApJ, 715, 1050

\bibitem[Delgado Mena et al.(2010)]{dm10} Delgado Mena, E., Israelian, G., Gonz\'alez Hern\'andez, J.I., Bond, J.C., Santos, N.C., Udry, S., \& Mayor, M. 2010, \apj, 725, 2349

\bibitem[Dressing \& Charbonneau(2013)]{dress13} Dressing, C.D. \& Charbonneau, D. 2013 ApJ 767, 95D

\bibitem[Elkins-Tanton \& Seager(2008)]{elk08} Elkins-Tanton, L.T. \& Seager, S. 2008, ApJ  688, 628

\bibitem[Gaidos(2013)]{gaidos13} Gaidos, E. 2013, ApJ 770, 90G

\bibitem[Gilli et al.(2006)]{gilli06} Gilli, G. , Israelian, G., Ecuvillion, A. Santos, N.C., \& Mayor, M. 2006 A\&A 449, 723

\bibitem[Graboske et al.(1973)]{graboske_1973_aa} Graboske, H. C., Dewitt, H. E., Grossman, A. S., and Cooper, M. S. 1973, \apj, 181, 457

\bibitem[Hayek et al.(2009)]{hayek09} Hayek, W., Wiesendahl, U., Christlieb, N., et al. 2009, A\&A, 504, 511

\bibitem[Hinkel et al.(2014)]{hinkel14} Hinkel, N.R., Timmes, F.X., Young, P.A., Pagano, M.D., Turnbull, M.C.. 2014, ApJ, 148, 54


\bibitem[Honda et al.(2004)]{honda04} Honda, S., Aoki, W., Kajino, T., et al. 2004, ApJS, 152, 113

\bibitem[Iglesias \& Rogers(1996)]{ir96} Iglesias, C.A., \& Rogers, F.J. 1996, ApJ, 464, 943

\bibitem[Iliadis et al.(2001)]{iliadis_2001_aa} Iliadis, C., D'Auria, J. M., Starrfield, S., Thompson, W. J., and Wiescher, M. 2001, \apjs,
134, 151

\bibitem[Kargel \& Lewis(1993)]{kargal93}Kargel, J. S., \& Lewis, J. S. 1993, Icarus, 105, 1

\bibitem[Kasting et al.(2013)]{kast13} Kasting, J.F., Kopparapu, R., Ramirrz, R.M., \& Harman, C. 2013 arXiv:1312.1328

\bibitem[Kasting, Whitmire, \& Reynolds(1993)]{kasting93} Kasting, J. F., Whitmire, D. P., and Reynolds, R. T. 1993, \icarus, 108, 108

\bibitem[Kopparapu et al.(2013a)]{kopp13a} Kopparapu, Ravi Kumar; Ramirez, Ramses; Kasting, James F.; Eymet, Vincent; Robinson, Tyler D.; Mahadevan, Suvrath; Terrien, Ryan C.; Domagal-Goldman, Shawn; Meadows, Victoria; Deshpande, Rohit  2013, \apj, 765, 131

\bibitem[Kopparapu et al.(2013b)]{kopp13b} Kopparapu, Ravi Kumar; Ramirez, Ramses; Kasting, James F.; Eymet, Vincent; Robinson, Tyler D.; Mahadevan, Suvrath; Terrien, Ryan C.; Domagal-Goldman, Shawn; Meadows, Victoria; Deshpande, Rohit 2013, \apj, 770, 82

\bibitem[Kuchner \& Seager(2005)]{kuchner05} Kuchner, M.J. \& Seager, S. arXiv:astro-ph/0504214

\bibitem[Kurucz(1993)]{kurucz93} Kurucz, R. 1993, ATLAS9 Stellar Atmosphere Programs and 2 kms$^{-1}$ grid. Kurucz CD-ROM No. 13 (Cambridge, Mass.: Smithsonian Astrophysical Observatory), 13

\bibitem[Lai et al.(2008)]{lai08} Lai, D. K., Bolte, M., Johnson, J. A., et al. 2008, ApJ, 681, 1524

\bibitem[Langanke \& Mart\`{i}nez-Pinedo(2000)]{LMP00} Langanke, K. and Mart\`{i}nez-Pinedo, G. 2000, Nucl. Phys. A, 673, 481

\bibitem[Bertan de Lis et al.(2015)]{lis15} Bertan de Lis, S., Delgado Mena, E., Adibekyan, V., Santos, N.C., Sousa, S.G.  arXiv:1501.05805

\bibitem[Madhusudhan et al.(2012)]{madhu}Madhusudhan,N.Ê Lee, K.ÊK.ÊM. andÊMousis, O. 2012 Astrophys.ÊJ.ÊLett., 759:L40.

\bibitem[Marcy \& Butler(1992)]{mb92} Marcy, G.W. \& Butler, R.P. 1992, PASP, 104, 270

\bibitem[Marcy et al.(2000)]{marcy00} Marcy, G., Butler, P.R., Fischer, D.A., \& Vogt, S.S. 2000, in ASP Conf. Ser. 213, Bioastronomy 99, ed. G. Lemarchand \& K. Meech (San Francisco, CA:ASP), 85

\bibitem[Meakin \& Arnett(2007)]{MA07_conv} Meakin, C.A. and Arnett, D. 2007, \apj, 667, 448

\bibitem[Milone et al.(2011)]{milone11} Milone, A., Sansom, A.E., Sanchez-Blazquez, P. 2011, MNRAS, 414, 1227

\bibitem[Neves et al.(2009)]{neves09} Neves, V., Santos, N.C., Sousa, S.G., Correia, A.C.M., \& Israelian, G. 2009, A\&A 497, 563

\bibitem[O\'Brien et al.(2006)]{obrien06} OÕBrien, D.P., Morbidelli, A., \& Levison, H.F. 2006, Icarus, 184, 39

\bibitem[Petigura et al.(2013)]{pet13} Petigura, E.A, Howard, A.W. \& Marcy, G.W. 2013 Proceedings of the National Academy of Sciences, 110, 48, pp. 19273-19278

\bibitem[Pijpers(2003)]{pij03a} Pijpers, F. P. 2003, \aap, 400, 241

\bibitem[Pijpers et al.(2003)]{pij03b} Pijpers, F. P., Teixeira, T. C., Garcia, P. J., Cunha, M. S., Monteiro, M. J. P. F. G., \& Christensen-Dalsgaard, J. 2003, \aap, 406, L15

\bibitem[Rauscher \& Thielemann(2000)]{rauscher_2000_aa} Rauscher, T. and Thielemann, F.-K. 2000, Atomic Data and Nuclear Data Tables 75, 1

\bibitem[Roederer et al.(2009)]{roederer09}Roederer, I.U., Kratz, K.L., Frebel, A., et al. 2009, ApJ, 698, 1963

\bibitem[Rogers \& Nayfonov(2002)]{rn02} Rogers, F.J. \& Nayfonov, A. 2002, ApJ, 576, 1064

\bibitem[Santos et al.(2001)]{santos01} Santos, N.C., Israelian, G., \& Mayor, M. 2001, A\&A, 373, 1019 

\bibitem[Santos et al.(2003)]{santos03} Santos, N.C., Israelian, G., Mayor, M., Rebolo, R., \& Udry, S. 2003, A\&A, 398,
363

\bibitem[Santos et al.(2005)]{santos05} Santos, N.C., Israelian, G., Mayor, M., Bento, J.P., Almeida, P.C., Sousa, S.G., Ecuvillon, A. A\&A 437, 1127

\bibitem[Sneden(1973)]{sneden73} Sneden, C. A. 1973, PhD thesis, University of Texas, Austin, Texas

\bibitem[Takeda(2007)]{take07} Takeda, Y. 2007 PASJ 59, 335

\bibitem[Tang \& Gai(2011)]{tang11} Tang, Y. K. \& Gai, N. 2011, \aap, 526, A35

\bibitem[Teixiera, et al.(2009)]{tex09} Teixeira, T.C., Kjeldsen, H., Bedding, T. R., Bouchy, F., Christensen-Dalsgaard, J., Cunha, M.S., Dall, T., Frandsen, S., Karoff, C., Monteiro, M.J.P.F.G., \& Pijpers, F. P. 2009, \aap, 494, 237

\bibitem[Thevenin(1998)]{thev98} Thevenin 1998yCat.3193....0T

\bibitem[Thoul, Bahcall, \& Loeb(1994)]{thoul_1994_aa} Thoul, A.A., Bahcall, J.N., and Loeb, A. 1994, \apj, 421, 828


\bibitem[Timmes \& Arnett(1999)]{timmes_1999_aa} Timmes, F. X. and Arnett, D. 1999, \apjs, 125, 277

\bibitem[Tuomi et al. (2013)]{tuo12} Tuomi, M. et al. 2013, A\&A, 551A, 79T

\bibitem[Umemoto et al.(2006)]{um06} K. Umemoto, R. M. Wentzcovitch, and P. B. Allen 2006 Science, 311:983-986

\bibitem[Unterborn et al.(2013)]{unt13} Unterborn, C.T., Kabbes, J.E., Pigott, J., Reaman, D., Panero, W.R. 2013 \apj, arXiv:1311.0024v3

\bibitem[Valenti \& Fischer(2005)]{valfis05} Valenti, J.A. \& Fischer, D.A. 2005, ApJS, 159, 141


\bibitem[Wiescher et al.(2006)]{wiescher_2006_aa} Wiescher, M., Azuma, R. E., Gasques, L., G\"{o}rres, J., Pignatari, M., and Simpson, E. 2006, Memorie della Societa Astronomica Italiana 77, 910

\bibitem[Wood et al.(2005)]{wood05} Wood, B.E.M.\"{u}ller, H.-R., Zank, G.P., Linsky, J.L., and Redfield, S. 2005, \apj, 628,143

\bibitem[Young \& Arnett(2005)]{young_2005_aa} Young, P.A. \& Arnett, D. 2005, \apj, 618, 908


\end{thebibliography}
\end{document}